\documentclass[12pt,preprint]{aastex}

\usepackage{ulem}

\usepackage{epsfig}

\begin{document}

\title{Comparisons of Cosmological MHD Galaxy Cluster Simulations to Radio Observations}

%% Notice placement of commas and superscripts and use of &
%% in the author list
\author{Hao Xu\altaffilmark{1,3}, 
Federica Govoni\altaffilmark{2},
Matteo Murgia\altaffilmark{2},
Hui Li\altaffilmark{1},  
David C. Collins\altaffilmark{1},
Michael L. Norman\altaffilmark{3},
Renyue Cen\altaffilmark{4},
Luigina Feretti\altaffilmark{5},
and Gabriele Giovannini\altaffilmark{5,6}
}
\altaffiltext{1}{Theoretical Division, Los Alamos National Laboratory, Los
  Alamos, NM 87545; hao\_xu@lanl.gov, hli@lanl.gov, dccollins@lanl.gov}
\altaffiltext{2}{INAF - Osservatorio Astronomico di Cagliari, Poggio dei Pini, Strada 54, 09012 Capoterra (CA), Italy; fgovoni@oa-cagliari.inaf.it, matteo@oa-cagliari.inaf.it}
\altaffiltext{3}{Center for Astrophysics and Space Science, University of California at San Diego, 9500 Gilman Drive, La Jolla, CA 92093; mlnorman@ucsd.edu }
\altaffiltext{4}{Department of Astrophysical Science, Princeton University, Princeton, NJ 08544; cen@astro.princeton.edu}
\altaffiltext{5}{INAF - Istituto di Radioastronomia, Via P.Gobetti 101, 40129 Bologna, Italy;  lferetti@ira.inaf.it, ggiovann@ira.inaf.it}
\altaffiltext{6}{Dipartimento di Astronomia, Universit\`a degli Studi di Bologna, Via Ranzani 1, 40127 Bologna, Italy}

\begin{abstract}
Radio observations of galaxy clusters show that there are $\mu$G magnetic fields permeating the intra-cluster medium (ICM), but it is hard to accurately constrain the strength and structure of the magnetic fields without the help of advanced computer simulations.
We present qualitative comparisons of synthetic VLA observations of simulated galaxy clusters to radio observations of Faraday Rotation
Measure (RM) and radio halos. The cluster formation is modeled using adaptive mesh refinement (AMR) magneto-hydrodynamic (MHD) simulations 
with the assumption that the initial magnetic fields are injected into the ICM by active galactic nuclei (AGNs) at high redshift. 
In addition to simulated clusters in \citet{Xu10, Xu11}, we present a new simulation with magnetic field injections from multiple AGNs. 
We find that the cluster with multiple injection sources is magnetized to a similar level as in previous simulations with a single AGN. 
The RM profiles from simulated clusters, both $|RM|$ and the dispersion of RM ($\sigma_{RM}$), are consistent at a first-order with the radial distribution from observations. The correlations between the $\sigma_{RM}$ and X-ray surface brightness from simulations are in a broad 
agreement with the observations, although there is an indication that the simulated clusters could be slightly over-dense and less magnetized with 
respect to those in the observed sample.
In addition, the simulated radio halos agree with the observed correlations
between the radio power versus the cluster X-ray luminosity and between the radio power versus the radio halo size. 
These studies show that the cluster wide magnetic fields that originate from AGNs and are then amplified by the ICM turbulence \citep{Xu10} match observations of magnetic fields in galaxy clusters.

\end{abstract}

%%%%%%%%%%%%%%%%%%%%%%%%%%%%%%%%%%%%%%%%%%%%%%%
\section{Introduction}
The detections of large-scale radio synchrotron emissions from galaxy clusters, called radio halos and relics \citep[see][]{Carilli02, Ferrari08, Giovannini09, Feretti12}, indicated that the intra-cluster medium (ICM) of galaxy clusters are permeated with cluster wide magnetic fields. The radio halos are diffusively extended over $\sim$ 1 Mpc, covering the whole clusters, while the radio relics are usually at the edges of the clusters as long filaments and could be related to recent cluster mergers \citep[e.g.][]{Bagchi06, Weeren10}.   By assuming that the magnetic energy is comparable to the total energy in relativistic electrons, one often deduces that the volume-averaged magnetic fields in the cluster halos are $\sim$ 0.1 to 1.0 $\mu$G and the total magnetic energy can be as high as 10$^{61}$ erg \citep{Feretti99} in the ICM of a galaxy cluster.  Recently, observations of radio halos have been used to study the structure of the cluster wide magnetic fields by comparing observations with mock halos from turbulent magnetic fields by construction \citep{Vacca10}.

The Faraday rotation measure (RM) is another important observational technique to study the properties of magnetic fields in galaxy clusters \citep[see][]{Carilli02}. Studies of RM by \citet{Eilek02, Taylor93, Colgate00} have suggested that the coherence scales of magnetic fields can range from a few kpc to a few hundred kpc.  Recently, the patchy RM maps were further studied to predict that the ICM magnetic fields have a Kolmogorov-like turbulent spectrum in the cores of some clusters \citep{Ensslin06, Kuchar11}.  
RM is widely used to estimate the strengths of the ICM magnetic fields from the cluster centers \citep[e.g.][]{Taylor93, Taylor01, Laing08, Guidetti10, Vacca12}
to the outer parts of clusters \citep[e.g.,][]{Clarke01, Murgia04, Govoni06, Guidetti08, Bonafede10}. Though there are large uncertainties in measuring the magnetic fields, these studies all suggest that the strengths of magnetic fields in the cluster centers range from 10s $\mu$G in the cool core clusters to a few  $\mu$G in the non-cool core clusters, and the magnetic field strengths drop gradually to  sub $\mu$G at the edges of clusters. In addition, \citet{Dolag01} showed that there is a power law correlation between two observables, the dispersions of RM ($\sigma_{RM}$) and the X-ray surface brightness (S$_{X}$), and suggested that the strengths of magnetic fields are related to the gas temperatures of the ICM. This $\sigma_{RM}$-S$_{X}$ correlation was further confirmed recently by \citet{Govoni10} with more observational data.

Though the ICM magnetic fields may not be dynamically important for the cluster formation as a whole \citep{Widrow02},  the presence of magnetic fields and their structure are important to many interesting astrophysics and plasma physics phenomena, e.g., anisotropic thermal conduction and acceleration and propagation of cosmic rays \citep[e.g.,][]{Parrish09, Brunetti07, Ensslin11}. Some of these processes, like thermal conduction, may play an important role in the dynamics and thermal equilibrium in cores of galaxy clusters \citep{Voit11}.

Since the formation of galaxy clusters and evolution of the ICM magnetic fields are highly non-linear, thus are too complicated to be studied analytically, numerical simulations have been playing an important part in studying the thermal properties of the ICM \citep[e.g.,][]{Bryan98,  Motl04, Nagai07, Vazza11}, and the ICM magnetic field evolution \citep[e.g.,][]{Dolag02, Dubois08, Xu09, Bonafede11} of galaxy clusters. Although the existence of cluster-wide magnetic fields is well accepted, their origin, which may be important to the evolution of magnetic fields during the course of cluster formation, is still unclear \citep{Widrow02, Dolag08}. MHD cluster formation simulations have been performed with different initial magnetic fields, including random or uniform fields from high redshifts \citep{Dolag02, Dubois08, Dubois09}, or from the outflows of normal galaxies \citep{Donnert08} or active galaxies \citep{Xu09}.  The cluster magnetic fields of all these simulations at low redshifts are roughly in agreement with each other, with $\mu$G in the cluster centers and decreasing with radius, and can produce some mock observations similar to available radio observations.  So the exact origins of magnetic fields may not be differentiated by the cluster magnetic fields at the current epoch. 

Large scale magnetized radio jets and lobes from AGNs serve as a very intriguing source of seed magnetic fields in clusters, because they could carry large amounts of magnetic energy and fluxes and distribute them to scales of hundreds kpc \citep{Burbidge59, Kronberg01, Croston05, McNamara07}.  The simulation presented in Xu et al (2009) suggests that cluster-wide $\mu$G magnetic fields can be produced from a single AGN that is further amplified by the small scale dynamo \citep{Brandenburg05}. Additional studies \citep{Xu10, Xu11} show that this mechanism is mostly insensitive to the injected energy and redshift, as well as cluster mass and formation history. The RM from these simulations show similar features of long filaments and small scale bands as in the observations from VLA. But to know how the AGN magnetic fields fit the real cluster magnetic fields, detailed comparisons between the synthetic observations of simulated clusters and the radio observations are needed. Such studies are necessary before applying these simulated magnetic fields to make predictions for the next generation of radio observations, such as EVLA, LOFAR, and SKA, or to the study of the particle acceleration by MHD turbulence in the ICM.

In this paper, we first present the results of a new simulation with magnetic field injections from multiple AGNs. In this simulation, the initial magnetic fields are from 30 AGNs in a volume that eventually forms a single cluster. This setup of injections is more realistic than previously simulations with only one AGN.  
We then compare the VLA filtered synthetic radio observations, including RM and radio halos from our high resolution MHD cosmological simulations to radio observations. 
To have a large sample of simulated clusters, the comparisons include the simulated clusters from our previous studies \citep{Xu10, Xu11}. We compare the radial profiles of absolute value of RM ($|RM|$) with results in \citet{Clarke01}, the $\sigma_{RM}$ with observations in \citet{Govoni10}, and the $\sigma_{RM}$ vs. S$_{X}$ relations with those in \citet{Govoni10}. For radio halos, we follow the method stated in \citet{Murgia04} and \citet{Murgia09} to generate synthetic radio halos by assuming energy equipartition between magnetic fields and non-thermal electrons. We then compare the statistical properties of the mock radio halos with the real ones.

The data processing and analysis of the simulations are performed using yt\footnote{http://yt.enzotools.org} \citep{Turk11} with additional modules for magnetic fields, and the FARADAY code \citep{Murgia04}.

This paper is structured as follows. We begin with a description of our new simulation in Section \ref{sec:simulation}. In Section \ref{sec:comparison} , we present the comparisons of  synthetic observations with radio observations, including RM and radio halos. Finally, we briefly discuss our results and conclusions in the last section. We refer to the Appendix for the details of the procedures we follow to generate the synthetic observations of RM and radio halos from the simulations.

%%%%%%%%%%%%%%%%%%%%%%%%%%%%%%%%%%%%%%%%%%%%%%%%%%%%%%

\section{Simulation of Galaxy Cluster Formation with Multi-AGN Magnetic Field Injections}
\label{sec:simulation}
\subsection{Numerical Method and Model}
\label{sec:model}

We first present our new self-consistent high-resolution cosmological MHD cluster formation simulation with magnetic fields initially from multiple AGNs. The simulation is performed using cosmological MHD code with adaptive mesh refinement (AMR) ENZO+MHD \citep{Collins09}. This code uses the AMR algorithms developed by \citet{Berger89} and \citet{Balsara01}, the MHD solver of \citet{Li08}, and the 
constrained transport (CT) method of \citet{Gardiner05}. The simulation here uses an adiabatic equation of state, with the ratio of specific heat, $\Gamma$ = 5/3, and does not include heating and cooling physics or chemical reactions, which are not important in this study.
 
The initial conditions of the simulation are generated at redshift $z=30$ from an \citet{Eisenstein99} power spectrum of density fluctuation in a $\Lambda$CDM universe with parameters $h=0.73$, $\Omega_{m}=0.27$, $\Omega_{b}=0.044$, $\Omega_{\Lambda}=0.73$, and $\sigma_{8}=0.77$. These parameters are close to the values from WMAP3 observations \citep{Spergel07}. The simulation volume is ($128$ $h^{-1}$Mpc)$^{3}$, and it uses a $128^3$ root grid and $2$ level nested static grids in the Lagrangian region where a cluster forms. This gives an effective root grid resolution of $512^3$ cells ($\sim$ 0.35 Mpc) and dark matter particle mass resolution of $1.34 \times 10^{9}M_{\odot}$, which is 8 times better than the dark matter resolution in previous simulations \citep{Xu10, Xu11}. In this paper, we only study the magnetic field injection and evolution in the central cluster out of  lots of galaxy clusters formed in the simulation domain. During the course of simulation, $7$ levels of refinements are allowed beyond the root grid, for a maximum spatial resolution of $7.8125$ $h^{-1}$ kpc, the same as the simulations in \citet{Xu10, Xu11}. The AMR is applied only in a volume of ($\sim$ 43 Mpc)$^3$ where the galaxy cluster forms near the center of the simulation domain. The AMR criteria in this simulation are the same as in \citet{Xu10}. During the cluster formation but before the magnetic fields are injected, the refinement is controlled by baryon and dark matter density. After magnetic field injections, all the regions where magnetic field strengths are higher than  5 $\times$ 10$^{-8}$ G are refined to the highest level, in addition to the density refinement.  
 
The magnetic fields are injected using the same method used in \citet{Xu08a, Xu09} as the original magnetic tower model proposed by \citet{Li06}.  We have the magnetic fields injected at redshift $z=2.5$ in the 30 most massive proto-clusters in the cluster forming region of $\sim$ (30 Mpc)$^3$, assuming that there is a super massive blackhole (SMBH) in each of these proto-clusters. Out of the 30 proto-clusters with injected magnetic fields, 26 finally form the major part of  the cluster, while the other four are still in the filaments. We assume that the magnetic fields are from $\sim$ 10$^{8}$ M$_\odot$ SMBHs with $\sim$ 1\% of outburst energy in magnetic form. So we put $\sim$ 1.3 $\times$ 10$^{59}$ erg magnetic energy into the IGM from each injection, and the total injected magnetic energy from all AGNs is about 4 $\times$ 10$^{60}$ erg. The number and masses of supermassive black holes used in this simulation are consistent with black hole mass density obtained by Sloan Digital Sky Survey \citep{Yu02}. We chose to inject the magnetic fields at the redshift $z=2.5$, which is close to the peak of comoving quasar number density \citep{Fan01, Fan06}. Previous studies \citep{Xu10} have shown that  the distribution of the ICM magnetic fields at low redshifts is not very sensitive to the exact injection redshifts and injected magnetic energies.

\subsection{Properties of the Galaxy Cluster and Distribution of Magnetic Fields}
\label{sec:result}

\begin{figure}
\begin{center}
\epsfig{file=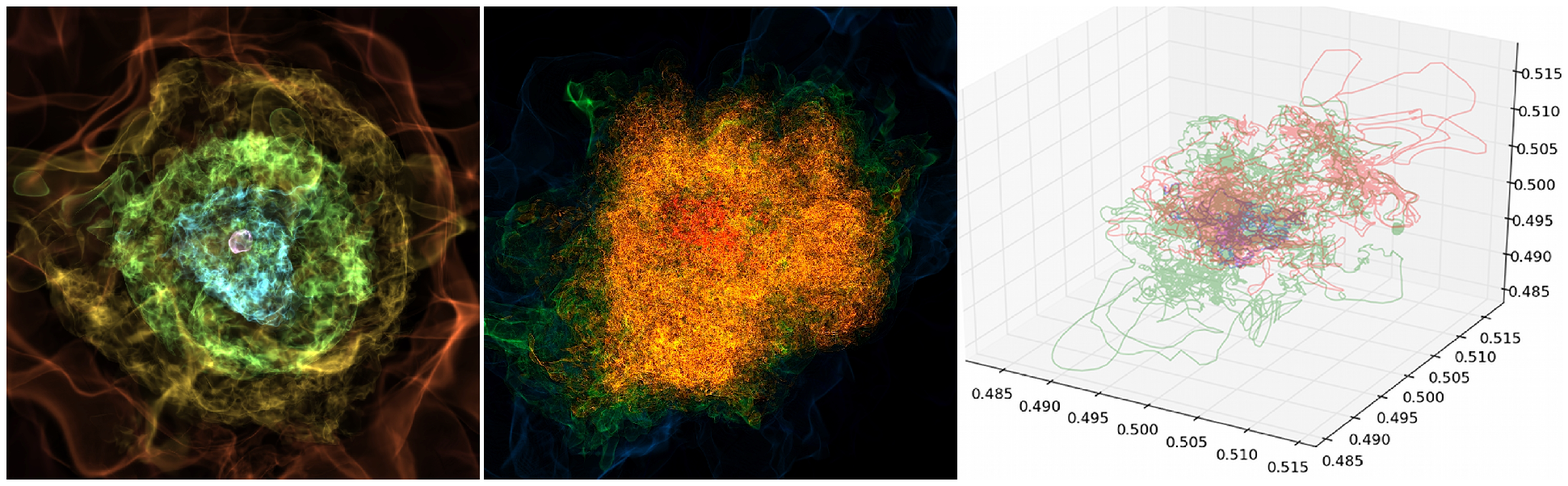,width=1.0\textwidth}
\end{center}
\caption{Isocontours of baryon density (left), magnetic field strength (middle), and magnetic field lines at $z=0$. 
The field of view is 6 Mpc.  The color ranges are 1 $\times$ 10$^{-31}$ (dark) to 3 $\times$ 10$^{-26}$ (bright) g cm$^{-3}$  
for density, and 1 $\times$ 10$^{-10}$ (blue) to 2 $\times$ 10$^{-6}$ (red) G
for magnetic fields. Different magnetic field lines are represented by different colors. 
 \label{fig:DEN_B}}
\end{figure}

Before we present the comparisons between observations and simulations, we first describe the properties and the magnetic field distribution of this simulated cluster with multi-AGN magnetic field injections. This simulated cluster is a massive cluster with its basic properties at redshift $z=0$ as follows:  R$_{virial}$ = 1.66 Mpc, M$_{virial}$(total) = 5.68 $\times$ 10$^{14}$ M$_{\odot}$, M$_{virial}$(gas) =  7.75 $\times$ 10$^{13}$ M$_{\odot}$, and T$_{virial}$ = 4.52 keV. The magnetic fields in this cluster are similarly distributed as in the clusters of single AGN injection simulations in \citet{Xu11}. In Figure \ref{fig:DEN_B}, we show the images of the isocontours of the baryon density and magnetic field strength, and a sample of magnetic field lines of this simulated galaxy cluster at redshift $z=0$. This cluster is disturbed by a big merger after redshift $z=0.25$ and is still unrelaxed at the current epoch,  so it looks irregular in shape. The isocontours of baryon density show shell structure visually with the highest baryon densities at the cluster center and then the densities decreasing monotonically with radius in all directions. But there is no such shell structure of the isocontours of magnetic field strength, showing that magnetic fields have no clear dependence on the distance to the cluster center (or the gas density) in 3-dimensional spatial distribution. This is likely because large amount of magnetic fields are generated by the dynamo process of turbulence in additional to the simple compression of ''frozen-in'' magnetic fields. The plot of magnetic field lines further shows that the magnetic fields are highly entangled as the magnetic field lines are stretched and twisted by the ICM turbulence,  while there are few field lines extended out of the cluster.

\begin{figure}
\begin{center}
\epsfig{file=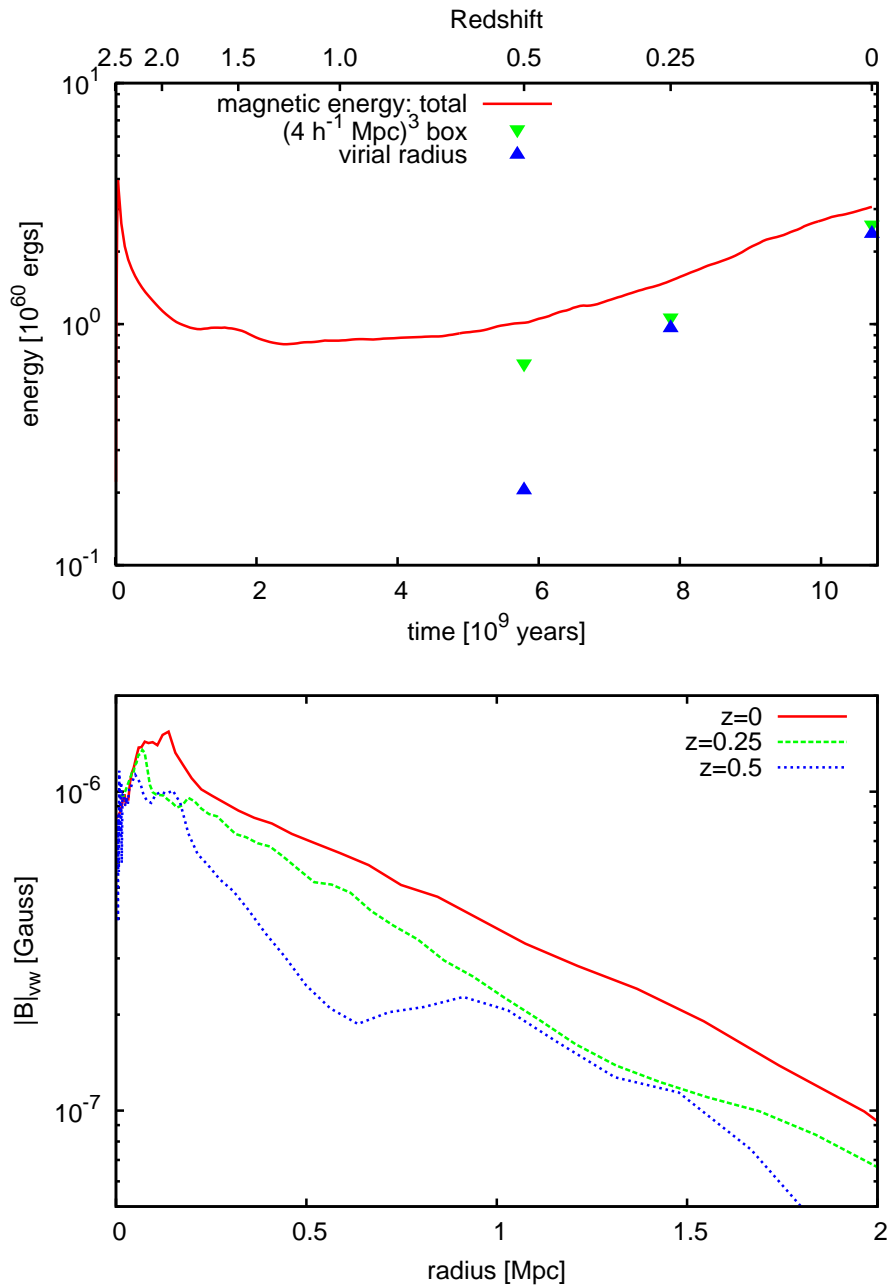,height=0.8\textheight} 
\end{center}
 \caption{ Top: Evolution of the total magnetic energy (solid line) in the simulation. The magnetic energies inside a box of (4 h$^{-1}$ Mpc)$^{3}$ comoving centered at the cluster center 
 and inside the virial radius of the cluster are marked at redshifts 
$z$ = 0.5, 0.25, and 0. At $z=0.5$, most of the magnetic fields (80\% of the total magnetic energy ) still reside in the incoming halos, and are out of the major cluster.   
 Bottom: The spherically averaged radial profiles of RMS magnetic field strength at redshifts $z$ = 0.5, 0.25, and 0.
 \label{fig:EB_B}}
\end{figure}

The magnetic energy evolution is shown in the top panel of Figure \ref{fig:EB_B}.  At redshift $z=0$, the magnetic energy inside the cluster virial radius is 2.37 $\times$ 10$^{60}$ erg, while the total magnetic energy in the simulation domain is 3.07 $\times$ 10$^{60}$ erg. So $\sim$ 77\%  of  the total magnetic energy is inside the galaxy cluster. The $\alpha$ parameter in Equation 1 of \citet{Xu11}, which measures the relation between magnetic energy and cluster mass, is 1.21, slightly bigger than those ($\sim$ 1) of the single AGN injection clusters in \citet{Xu11}. So this cluster, though having much more initial magnetic fields, still follows the E$_{M}$ $\propto$ M$_{virial}^2$ scaling found in single AGN model simulations. We also show the spherically averaged radial profiles of the magnetic field strengths at $z=0.5$, $0.25$,  and $0.0$ in the bottom panel of  Figure \ref{fig:EB_B}.  Though the 3-dimensional dependence of magnetic field strength on radius is weak, the averaged magnetic field strengths drop with radius monotonically (see \citet{Xu11} for a detailed discussion on the dependence of field strength on radius and gas density). The mean magnetic field strengths at $z=0$ are just above 1 $\mu$G in the cluster center and decrease slowly with radius to about 0.2 $\mu$G at the virial radius. The magnetic field strengths have little evolution between $z=0.5$ and $z=0$, except some bumps related to mergers at $z=0.5$.  Based on the mean magnetic field profiles of clusters in Xu et al. (2011), one would expect an unrelaxed cluster such as this to have an irregular radial profile.  However, the radial profiles here are smooth, most likely due to the fairly regular distribution of the magnetic injection sites. In addition, these profiles show that the magnetic field strength does not peak at the cluster center. Since magnetic fields are significantly amplified by the ICM turbulence \citep{Xu10},  the density and magnetic field strength peaks do not necessarily coincide. 

%%%%%%%%%%%%%%%%%%%%%%%%%%%%%%%%%%%%%%%%%%%%%%%%%%%%%%%%%%%%

\section{Comparisons of Simulated Results with Radio Observations}
\label{sec:comparison}
To see how well the magnetic fields in our simulated clusters fit the observations, we compare the synthetic radio observations from our simulations with observations of RM and radio halos. In addition to our multi-AGN run, we also include a set of twelve galaxy clusters with a wide range of sizes and formation histories with single AGN magnetic field injection from \citet{Xu11}.

\subsection{Simulated and Synthetic Faraday Rotation measure Images}

\begin{figure}
\begin{center}
\epsfig{file=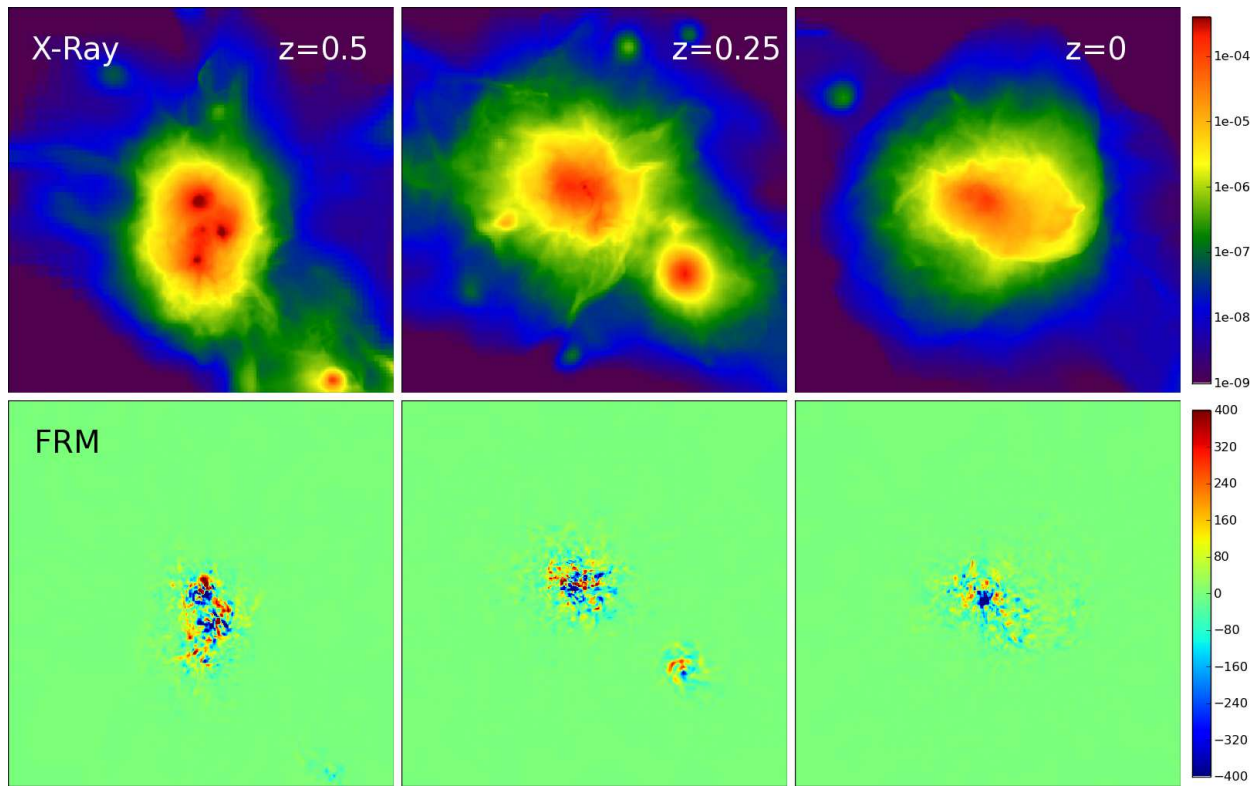,width=1.0\textwidth}
\end{center}
\caption{Top panel: Simulated X-ray emissions observed from the y direction in the 0.1 to 2.4 keV band of the cluster at redshifts $z=0.5,~0.25$, and $0$. Each image covers an area of $4$ h$^{-1}$ Mpc  
  $\times$ $4$ h$^{-1}$ Mpc comoving, while the X-rays are calculated from boxes of (4 h$^{-1}$ Mpc)$^3$.  The color range is from 1 $\times$ 10$^{-9}$
  (blue) to 4 $\times$ 10$^{-4}$ (red) erg s$^{-1}$ cm$^{-2}$. 
Bottom panel: Simulated Faraday rotation measure of the cluster by integrating
  through the cluster on the y direction at $z=0.5,~0.25$, and $0$. 
They are calculated from the same boxes as for the X-ray.  
 The color range is from $-400$ (blue) to $400$ (red) rad m$^{-2}$. 
\label{fig:XrayRM}}
\end{figure}

The simulated X-ray and RM images for the  multi-AGN run at redshift $z=0.5$, $0.25$, and $0$ are shown in Figure \ref{fig:XrayRM}.
The X-ray images are calculated within a box of (4 h$^{-1}$ Mpc)$^3$ (comoving)
in the band of 0.1 to 2.4 keV using yt. The simulated  RM maps are calculated within the same box assuming that a polarized synchrotron source is behind\footnote{Note that in this way we are overestimating the RM by approximately a factor of $\sqrt{2}$ with respect to the case in which the source is located half-way at the center of the cluster.} the whole volume
\begin{equation}
RM_{sim}=812 \int_{0}^{L} n_{e} B_{y} dy ~~~\rm (rad/m^2), 
\label{rmsim}
\end{equation}
where $n_{e}$ is the electron number density in cm$^{-3}$, $B_{y}$ the magnetic field component along the line-of-sight in $\mu G$, and 
$L$ is the size of the computational box in kpc. There is clear correlation between the X-ray luminosity and the RM, as the significant RM ($>$ 100 rad m$^{-2}$) only appear within bright X-ray ($>$ 1 $\times$ 10$^{-5}$ erg cm$^{-2}$ s$^{-1}$) regions. The RM patterns are patchy and show both large and small scale structures, as expected
 given the turbulent state of the simulated magnetic fields in the ICM.

In the following, we compare the simulated Faraday rotation from the multi-AGN injection simulations, as well as the results from \citet{Xu11}, with observations from \citet{Clarke01} and \citet{Govoni10}.  \citet{Clarke01}  report the observed mean RM for radio sources at different impact parameters from the centers of a sample of nearby galaxy clusters.  \citet{Govoni10} report the dispersion of the Faraday rotation ($\sigma_{RM}$) calculated from the distribution of all pixel values of spatially resolved RM images of a sample of radio sources in (or behind) nearby galaxy clusters. Both these indicators derive ultimately from Eq.\ref{rmsim}, but according to \cite{Murgia04} the two quantities trace the magnetic field on different scales. For an isotropic turbulence, the $|RM|$ offset from zero would be due to ICM magnetic field fluctuations on
 scales larger than the projected radio source's size at the distance of the cluster. The $\sigma_{RM}$ instead would be due to magnetic field fluctuations on scales smaller than the projected radio source's size. Indeed, the ratio of  $|RM|$/ $\sigma_{RM}$ is directly 
 related to the intra-cluster magnetic field power spectrum. In general the $|RM|$ from point-to-point in a cluster 
is more scattered since it involves a smaller number of magnetic field fluctuations along the line-of-sight and it is sensitive 
to the foreground Faraday rotation of our Galaxy, which must be subtracted from the data. The $\sigma_{RM}$ is statistically 
more stable, since it depends on a large number of small scale fluctuations, and it is mostly insensitive to unrelated 
foreground Faraday screens, but it requires high quality RM images to be measured. Usually, only a small number of sources are
 observed per cluster, thus only a few limited patches of the magneto-ionic medium are accessible through the observed RM images.
Indeed, for a proper comparison with observations, the simulated RM images have
been filtered as typical VLA polarization observations, as described 
in Appendix A. We then calculate the averaged radial profiles of the absolute value of $|RM|$ and the dispersion $\sigma_{RM}$. These radial profiles have been obtained from the average of the scatter of the RM statistics computed over boxes of $100 \times 100$ kpc$^2$  which is the typical size of observed RM images of cluster radio sources (see Appendix for details). In the following, we refer to the filtered RM images as the ``synthetic'' RM images, to distinguish them from the ``simulated'' RM images shown in the bottom panels of Fig.\,\ref{fig:XrayRM}. In the following we perform a qualitative comparison of the synthetic RM with the observations with 
the aim to determine whether the magnetic fields in our simulated clusters are able to provide, at least at the qualitative level, a consistent description of the observed magnitude and trends of the RM signal. A detailed quantitative analysis would require a much larger statistics of both simulated and observed clusters and it is beyond the scope of this work.

\subsubsection{Comparison with Observed $|RM|$ Profiles}

We first compare the radial profiles of $|RM|$ of the synthetic RM maps from the multi-AGN injection simulations, as well as the results from \citet{Xu11}, with observations from \citet{Clarke01}. 

\begin{figure}
\begin{center}
\epsfig{file=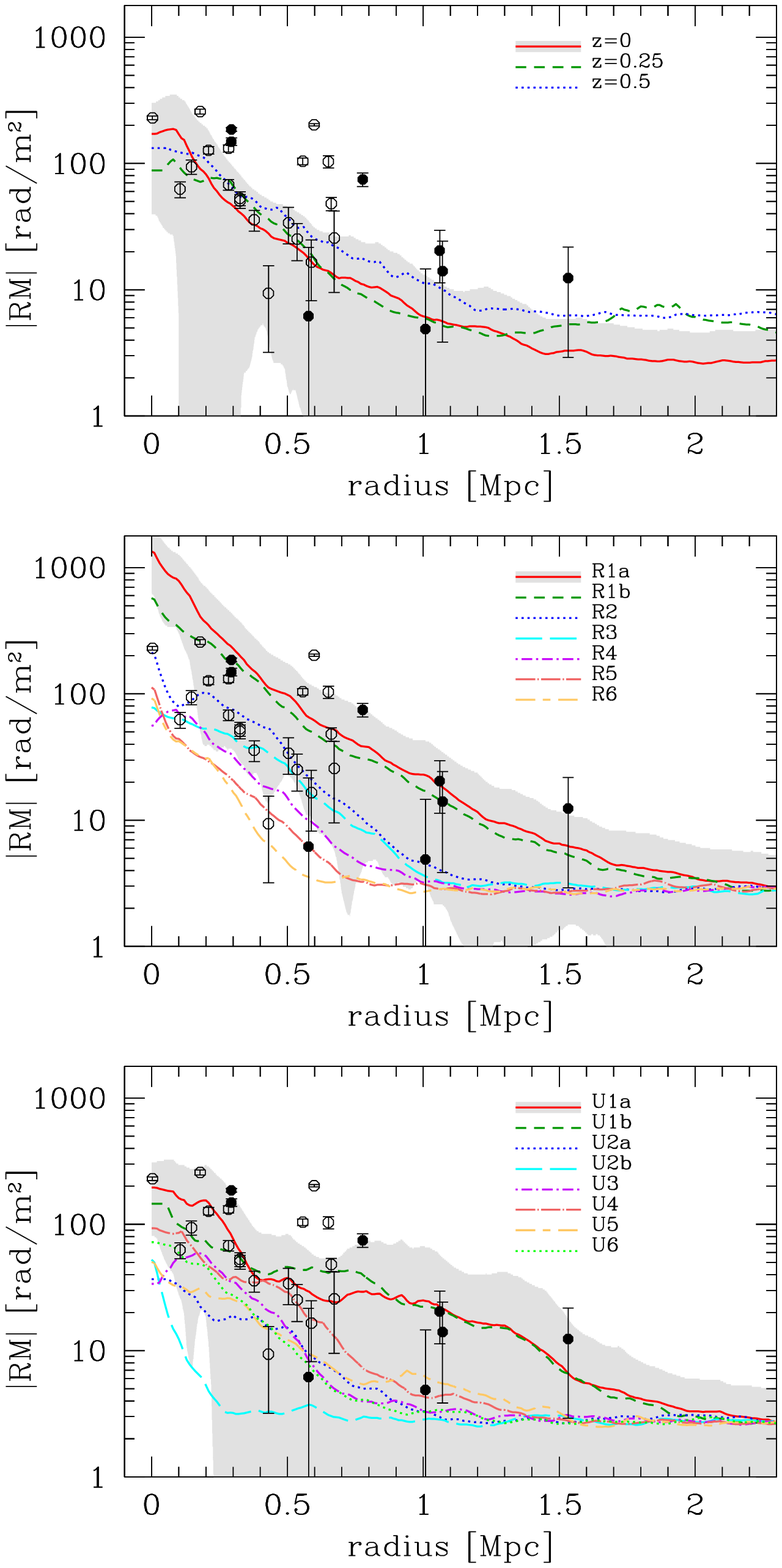,height=0.8\textheight}
\end{center}
 \caption{Azimuthally averaged radial profiles of $|RM|$. The panels, from top to bottom, show the 
 results of the multi-AGN run at different redshifts, a set of relaxed clusters, and a set of unrelaxed clusters, respectively.   
 The observations from \citet{Clarke01} are over-plotted 
in two groups based on whether the clusters are hotter (filled dots) 
or cooler (empty dots) than 5 keV. The shaded regions represent the rms scatter around the average profiles for multi-AGN, R1a, and U1a runs, respectively.
   \label{fig:RMy}}
\end{figure}

We plot the radial profiles of $|RM|$ of the multi-AGN case at various redshifts in the top panel in Figure \ref{fig:RMy}, where  the lines represent the average of the $|RM|$ of all the boxes
at a given radius from the cluster center. It must be noted that, due to the fluctuations of the intra-cluster magnetic field on scales larger than 100 kpc (i.e. the size of the boxes in which 
the RM statistics is calculated), the synthetic profiles are characterized by a strong intrinsic scatter, with a dispersion of almost one order of magnitude around the average. We show the rms scatter as a shaded region for the RM profile at redshift $z=0$. 
The $|RM|$ profiles at $z=0.25$ and $z=0.5$ have a similar scatter (not shown for graphical simplification).

Then we show the profiles from the relaxed and unrelaxed clusters\footnote{Note that we name ``relaxed'' those simulated clusters which are in the post merger stage. These systems should not be confused with the relaxed cool-core clusters often referred in the literature.} in \citet{Xu11} at $z=0$ in the middle and bottom panels, respectively. 
These clusters are called by the same names as in \citet{Xu11}. There are two profiles of cluster R1 shown, designated as R1a and R1b, as the simulations A and B in \citet{Xu10}, which  are from the same simulated cluster with two different amounts of injected magnetic fields. All the simulated clusters will be named in this way throughout the paper. The temperatures of these clusters are 7.65, 5.9, 4.14, 3.04, 2.16, and 1.41 keV, for the relaxed clusters from R1 to R6, and 10.26, 4.84, 4.78, 4.59, 4.34, and 3.43 keV, for the unrelaxed clusters from U1 to U6. 
The shaded regions in middle and bottom panels correspond to the intrinsic scatter of the R1a and U1a simulations, respectively.

The results of observations from \citet{Clarke01} are over plotted. The data are from different clusters, since it is still not possible to measure a large number of RM at different radii in a single cluster due to the limited sensitivity of current polarization observations.
Two different symbols are used to show whether the cluster temperatures are hotter (filled dots) or cooler (empty dots) than 5 keV. The temperatures of the observed clusters are taken from literature \citep{Proust03, Chen07, Bogdan11}.

The observed $|RM|$ from \citet{Clarke01} have a clear trend that the absolute values of RM drop gradually with increasing radius, while $RM$ have typical values of 100s rad m$^{-2}$ near the cluster centers. The synthetic $|RM|$ profiles from all simulations clearly show similar trends, but have some different features, likely due to the differences in cluster properties and in the cluster formation histories. The $|RM|$ profiles from the multi-AGN simulation are smooth and match the small $|RM|$ from observations. It is likely because this simulated cluster is not very large, just a little cooler than 5 keV, and its magnetic volume filling is high. Furthermore, the $|RM|$ profiles, similar to the magnetic field strength profiles, have little evolution between $z=0.5$ and $z=0$. 
The profiles from single AGN injected clusters clearly show a positive trend between the cluster mass (temperature) and the $|RM|$, and the profiles from large clusters cover those large observed $|RM|$. The $|RM|$ from relaxed clusters are usually larger than the $|RM|$ from unrelaxed clusters in similar size. 

We note once again that the $|RM|$ is an intrinsically scattered indicator. The nature of the scatter is not due to instrumental errors but rather represents 
an intrinsic characteristic of the intra-cluster magnetic fields, as illustrated by the shaded regions shown in Figure \ref{fig:RMy} and by the spread
 of observed data points as well. The scatter of the $|RM|$ inside a single cluster is comparable to the difference between the average $|RM|$ profiles among different clusters. It is then obvious that a detailed comparison would required a much larger statistics of both simulated and observed galaxy clusters.
Nevertheless, from this first comparison, we can conclude that most of the simulated $|RM|$ profiles lie within the range of values spanned by the data, with
 an overall agreement which is within the expected intrinsic scatter. There are, however, a few exceptions like simulations 
R6 and U2b whose $|RM|$ profiles are well below the data. On other hand, this is not surprising since these are among the least magnetized systems in 
the sample of simulated clusters described in \citet{Xu11}.

 \subsubsection{Comparison with Observed $\sigma_{RM}$ Profiles}

\begin{figure}
\begin{center}
\epsfig{file=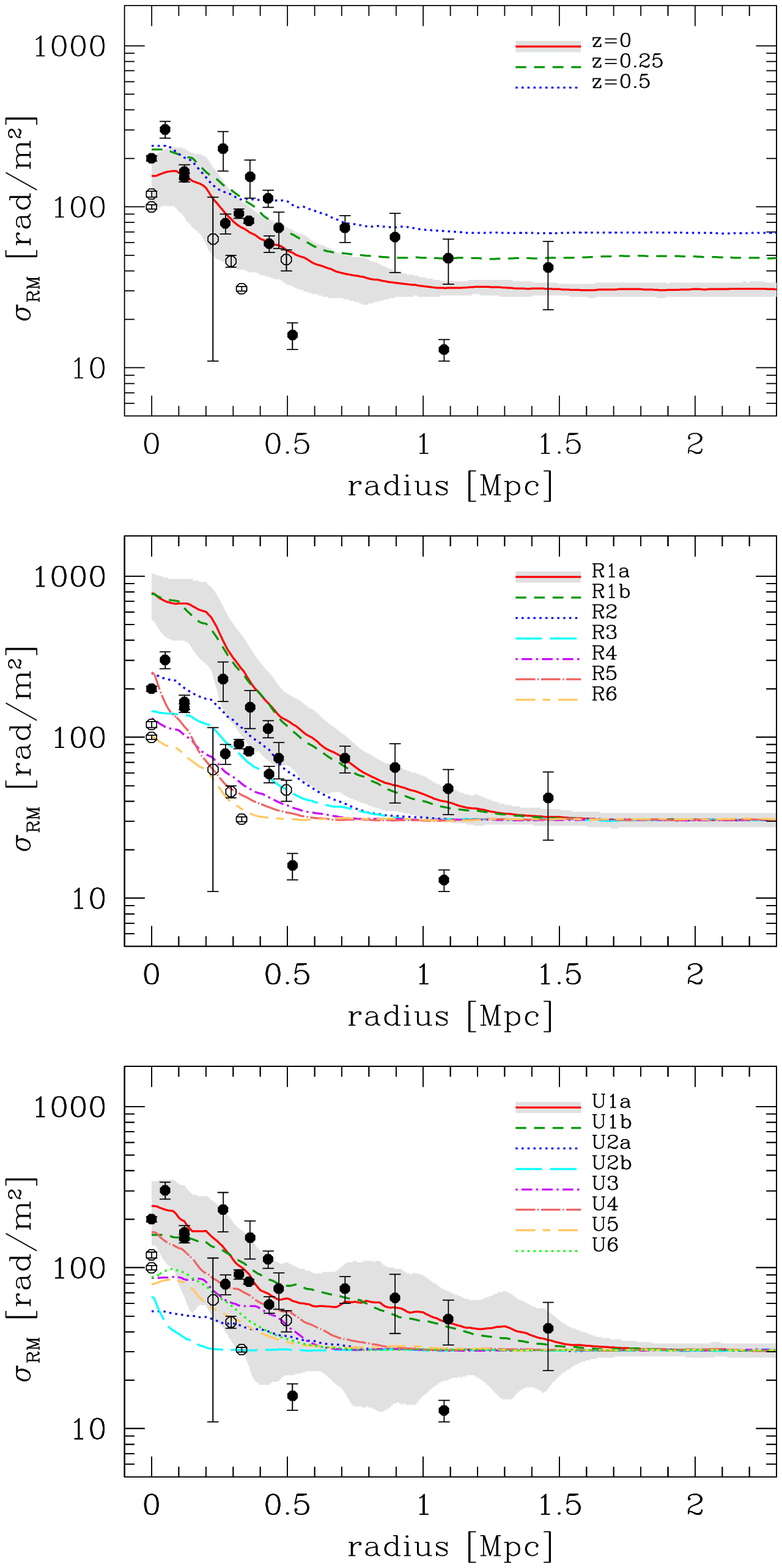,height=0.8\textheight}
\end{center}
 \caption{Radial profiles of the dispersion of RM distribution.
%calculated in concentric annuli. 
The panels, from top to bottom, show the results of the multi-AGN run at different redshifts, a set of relaxed clusters, and a set of unrelaxed clusters, respectively. The observational data from \citet{Govoni10} are over-plotted
in two groups based on whether the clusters are hotter (filled dots) 
or cooler (empty dots) than 5 keV. The shaded regions represent the rms scatter around the average profiles for multi-AGN, R1a, and U1a runs, respectively.
   \label{fig:sigmaRMy}}
\end{figure}

We plot the radial profiles of $\sigma_{RM}$ of the multi-AGN simulation at redshifts $z=0.5$, $0.25$,  and $0$ in the top panel of Figure \ref{fig:sigmaRMy}, while the profiles of relaxed clusters and unrelaxed clusters at $z=0$ in \citet{Xu11} are shown in the middle and bottom panels, respectively. The results from observations \citep{Govoni10} are over plotted, and are represented 
by different symbols in the cluster temperature, 
as listed in the table 5 of \citet{Govoni10}. As in Figure 4, we show, as shaded regions, the intrinsic scatter of the $\sigma_{RM}$ profiles for the multi-AGN simulation at redshift $z=0$ and for the R1a and U1a simulations. 

The $\sigma_{RM}$ have similar radial profiles as $|RM|$, but data and synthetic profiles 
(as illustrated by the shaded regions) are much less dispersed. The $\sigma_{RM}$ near the cluster centers are of the order of a few 100 $rad/m^2$ and gradually drop
 with radius. The flattening seen at large radii is an instrumental effect: the minimum $\sigma_{RM}$ that can be measured is limited by the noise in the observed 
RM images and we account for this effect in the synthetic RM images (see Appendix A).

The synthetic $\sigma_{RM}$ profiles are generally within the range spanned by the data.
The $\sigma_{RM}$ profiles for unrelaxed clusters are bumpy and the scatter is somewhat larger, 
probably as a result of recent mergers occurred in these systems.
In addition, the synthetic profiles reproduce the statistical trend observed in the data between the $\sigma_{RM}$ and the cluster temperature \citep{Govoni10}. Namely, the more massive, hotter galaxy clusters are also characterized by higher 
values of $\sigma_{RM}$.  Indeed, simulations expect that hotter clusters to be also more magnetized \citep[see][]{Xu11} and, as a consequence, they should present also higher $\sigma_{RM}$. 
However, as pointed out by \citet{Govoni10}, hotter clusters are also generally denser, at least in their sample, and thus it is not easy to determine which is the dominant effect between magnetic field strength and gas density. This holds in general for all simulated  $\sigma_{RM}$ profiles.
In fact, $\sigma_{RM}$ basically scales as a function of gas density and
 magnetic field  strength as
\begin{equation}
\sigma_{RM}\propto 812 \left[\Lambda_{c} \int_0^L (n_{e} B_{y})^2 dy \right]^{0.5} ,
\label{sigmarm}
\end{equation}
where $\Lambda_{c}$ is the auto-correlation length of the magnetic field fluctuations \citep[see e.g.][]{Murgia04}. The value of  $\sigma_{RM}$ depends directly on the product of $B$ and 
 $n_{e}$. From Figure \ref{fig:sigmaRMy} we show that
  simulations reproduce quite well the observed radial profiles of $\sigma_{RM}$, but we are not able to 
tell if this is due to the right combination of $B$ and $n_{e}$.

\subsubsection{Comparison with Observed $\sigma_{RM}$ -- S$_{X}$ Relation}
 
A possibility to break the degeneracy between magnetic field strength and gas density is to investigate the relationship between $\sigma_{RM}$ and the X-ray surface brightness
 S$_{X}$. Indeed, \citet{Dolag01} and \citet{Govoni10} showed that there is a positive correlation between these quantities. In fact, 

\begin{equation}
S_{X}\propto \int_0^L n_{e}^2 \sqrt{T} dy.
\label{sx}
\end{equation}
Thus, by comparing $\sigma_{RM}$ and S$_{X}$ we can try to reduce the degeneracy between $n_{e}$ and $B$ in Eq.\,\ref{sigmarm}.

\begin{figure}
\begin{center}
\epsfig{file=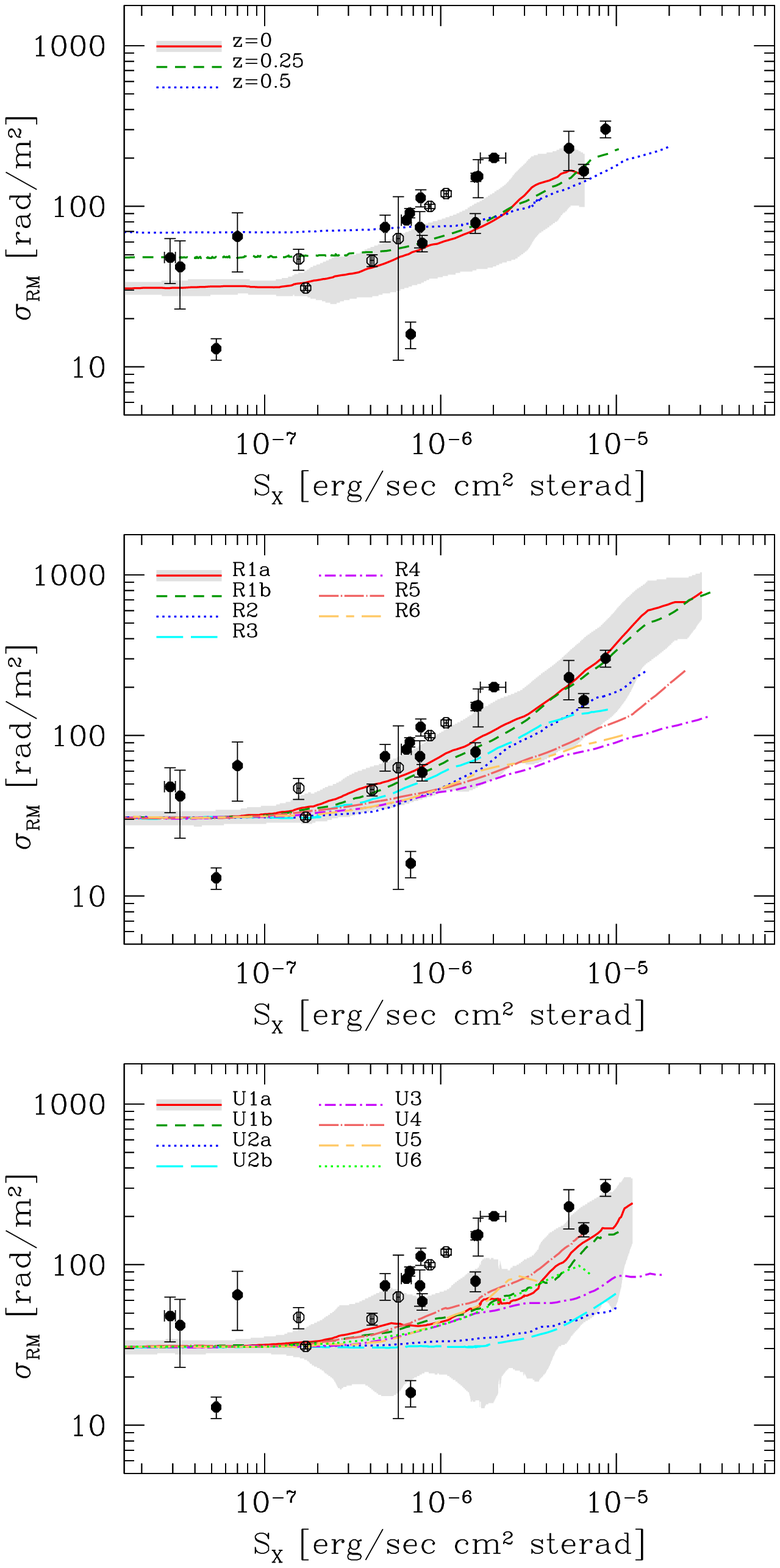,height=0.8\textheight} 
\end{center}
 \caption{Dispersion of RM distribution as a function of the X-ray surface brightness. The panels, from top to bottom, show the simulated  
 results of the multi-AGN run at different redshifts, a set of relaxed clusters, and a set of unrelaxed clusters, respectively.
The observational data from \citet{Govoni10} are over-plotted
in two groups based on whether the clusters are hotter (filled dots) 
or cooler (empty dots) than 5 keV. The shaded regions represent the rms scatter around the average profiles for multi-AGN, R1a, and U1a runs, respectively.
   \label{fig:xray_rm}}
\end{figure}

We present the $\sigma_{RM}$ -- S$_{X}$ relations from our simulations and the observations from \citet{Govoni10} in Figure \ref{fig:xray_rm}. The results from the multi-AGN simulation at redshifts $z=0.5$, $0.25$, and $0$ are plotted in the top panel, while the results from relaxed and unrelaxed clusters from \citet{Xu11} at redshift $z=0$ are shown in the middle and bottom panels, respectively.  The $\sigma_{RM}$ and the corresponding X-ray surface brightnesses have been computed consistently from the same regions of
 $100\times 100$ kpc$^2$. No redshift corrections of $\sigma_{RM}$ and S$_{X}$ had been done, since all the observations are at low redshifts ($<$ 0.1) and the corrections are not significant.  As in Figure 4 and 5, the shaded regions represent the intrinsic scatter of $\sigma_{RM}$ for the multi-AGN, R1a, and U1a simulated clusters.

We note that the simulated profiles show the same trend as data, in particular the slope 
of the simulated  $\sigma_{RM}$ -- S$_{X}$ relation is close to the observed one with a clear power-law regime at 
high $\sigma_{RM}$ and  S$_{X}$. At very low $\sigma_{RM}$ and S$_{X}$ the synthetic correlations flatten because the cluster RM signal fall below the detection threshold of the RM observations, which is about 30 rad/m$^2$. We recall that this is one of the most relevant features introduced by the instrumental filtering (see Appendix A). For the unfiltered simulated clusters (not shown), the power law regime of the $\sigma_{RM}$ -- S$_{X}$ relation extends for more than three order of magnitudes down to very low X-ray surface brightness and RM.

Although the synthetic profiles show a slope consistent with the observed trend, they appear to be 
  systematically shifted toward high S$_{X}$. This suggests that the simulated clusters could be slightly over-dense or under magnetized than the observed ones. Since S$_X$ also increases with the square root of T, it could be also possible that simulated clusters are systematically hotter than the observed ones. However, we checked that the distributions of global temperature for simulated and observed clusters are quite similar. Indeed, the average temperature for simulated cluster is of 4.7 keV which is even lower than the average temperature of 6.0 keV found for the galaxy clusters in Govoni et al (2010). This shift is more evident for the relaxed and unrelaxed simulated clusters with a single AGN injection, as the under magnetization is likely. For example, the radial profile of simulated cluster R1a reproduces the upper bound of data in Figure \ref{fig:sigmaRMy}
 but lies just in the middle of the  $\sigma_{RM}$ -- S$_{X}$ relation and the simulated S$_{X}$ peak exceeds that of data by roughly a factor of four.

In summary, the comparison of Figures \ref{fig:sigmaRMy} and \ref{fig:xray_rm} suggests that simulated clusters should by slightly 
 less dense and more magnetized to fit those in the \citet{Govoni10} sample. If simulated  magnetic fields were a factor
 of two stronger while gas densities a factor of two lower, $\sigma_{RM}$ 
 would remain unchanged but the expected $\sigma_{RM}$ -- S$_{X}$ relations would shift to the left by a factor of four.  The multi-AGN injection simulation seems to simultaneously provide a consistent description to both the  $\sigma_{RM}$ radial profiles and the $\sigma_{RM}$ -- S$_{X}$ relation, suggesting a better combination of $B$ and $n_e$ for this cluster. Again, there is no evolution of these relations from $z=0.5$ to $z=0$ for the multi-AGN injection model. It maybe because this cluster is well magnetized during all this time.

\subsection{\textbf{Simulated and Synthetic Radio Halo Images}}

The comparison between observed and synthetic radio halo images
is another tool to investigate magnetic fields in clusters of galaxies \citep{Murgia04, Govoni06, 
Vacca10}. This approach consists of simulations of radio halo images obtained by illuminating cluster magnetic fields models
 with a population of relativistic electrons.

In the specific case considered in this work, the radio halo modeling simulations have been performed at 1.4 GHz with a bandwidth of 25 MHz. 
At each point of the computation grid, we calculate the synchrotron emissivity by convolving 
the emission spectrum of a single relativistic electron with 
the particle energy distribution of an isotropic population of
relativistic electrons whose distribution follows
$N(\gamma,\theta)=K_{\rm 0}\gamma^{-\delta}(\sin\theta)/2 $,
where $\gamma$ is the electron's Lorentz factor, while $\theta$ is the pitch 
angle between the electron's velocity and the local direction of the 
magnetic field.
The energy density of the relativistic electrons is
$u_{\rm el}=m_ec^2\int_{\gamma_{\rm min}}^{\gamma_{\rm max}}N(\gamma)\gamma d\gamma$, 
where $\gamma_{\rm min}$ and  $\gamma_{\rm max}$ 
are the low and high energy cut offs of the energy spectrum, respectively.
In our modeling we assume equipartition between the magnetic field 
energy density $u_{\rm B}=B^2/8\pi$ and $u_{\rm el}$ at every 
point in the cluster, therefore both energy densities have the 
same radial decrease. 
Radio halos have a typical spectral index $\alpha=1.3$, 
thus we adopt an electron energy spectral index $\delta=2 \alpha +1=3.6$.
The low-energy cutoff $\gamma_{\rm min}$ and normalization $K_{\rm 0}$  
are adjusted to guarantee $u_{\rm el}= u_{\rm B}$, in practice 
we fixed $\gamma_{\rm min}$=300 and let $K_{\rm 0}$ vary. The energy losses of the relativistic electrons in the ICM reach a minimum for this value of the Lorentz factor \citep{Sarazin99}.
For the steep power law energy spectrum we considered, the exact value of $\gamma_{max}$ has only a marginal impact on 
the modeling, we adopted a fixed value of $\gamma_{max}=1.5\times 10^{4}$ for consistency with the radial steepening 
observed in the spectral index images of some radio halos \citep{Feretti12}. Most observations have failed to detect polarization in radio halos,
and thus in the present paper we focus on the analysis of the total intensity emission. Nevertheless, the investigation of the polarized
intensity of the simulated radio halos is important and will be the subject of a next publication.

We produced radio halo images for the simulated cluster (at redshift z=0)
with multi-AGN injections,
as well as for the simulated clusters by \citet{Xu11}: 
R1a, R1b, R2, R3, R4, R5, R6, U1a, U1b, U2a, U2b, U3, U4, U5, U6.
For a proper comparison with observations, the images have
been filtered as typical VLA radio halo observations, as described 
in Appendix B.
For each cluster we obtained a synthetic radio image 
with an angular resolution of 50$''$ FWHM and a noise level 
of 0.1 mJy/beam (1$\sigma$).
We found that in multi-AGN, R1a, R1b, R2, R3, U1a, U1b, and U4 the radio signal
is above 3$\sigma$ and the radio emission is strong enough
to be detected in a typical VLA observation.
In all the other simulations the signal is below the 3$\sigma$ limit 
and only an upper limit to the flux density can be given.
Following the same approach adopted for the RM simulations, we compared 
the synthetic radio halos with the observations. In particular,
we focus on the offset between the radio and the X-ray peaks, 
on the correlation between the global radio luminosity at 1.4 GHz and the X-ray luminosity ($P_{\rm 1.4 GHz}$ -- $L_{X}$), and
on the correlation between the Largest Linear Size ($LLS$), as measured from the $3\sigma$ isophote
on the synthetic images, and the radio power. These properties are derived from the synthetic halo images
 and are compared to the data taken from the literature \citep{Feretti12, Govoni12}.
The evolution of the relativistic electron component is not treated explicitly in our simulations and hence
we cannot predict the formation of radio halos. Our aim here is to check whether, given a reasonable energy 
spectrum for the synchrotron electrons, the simulated magnetic field produces radio halos whose global properties are
 in line with the  observations.

\subsubsection{Comparison with the Observed Radio to X-ray Offsets}
\begin{figure}
\begin{center}
\epsfig{file=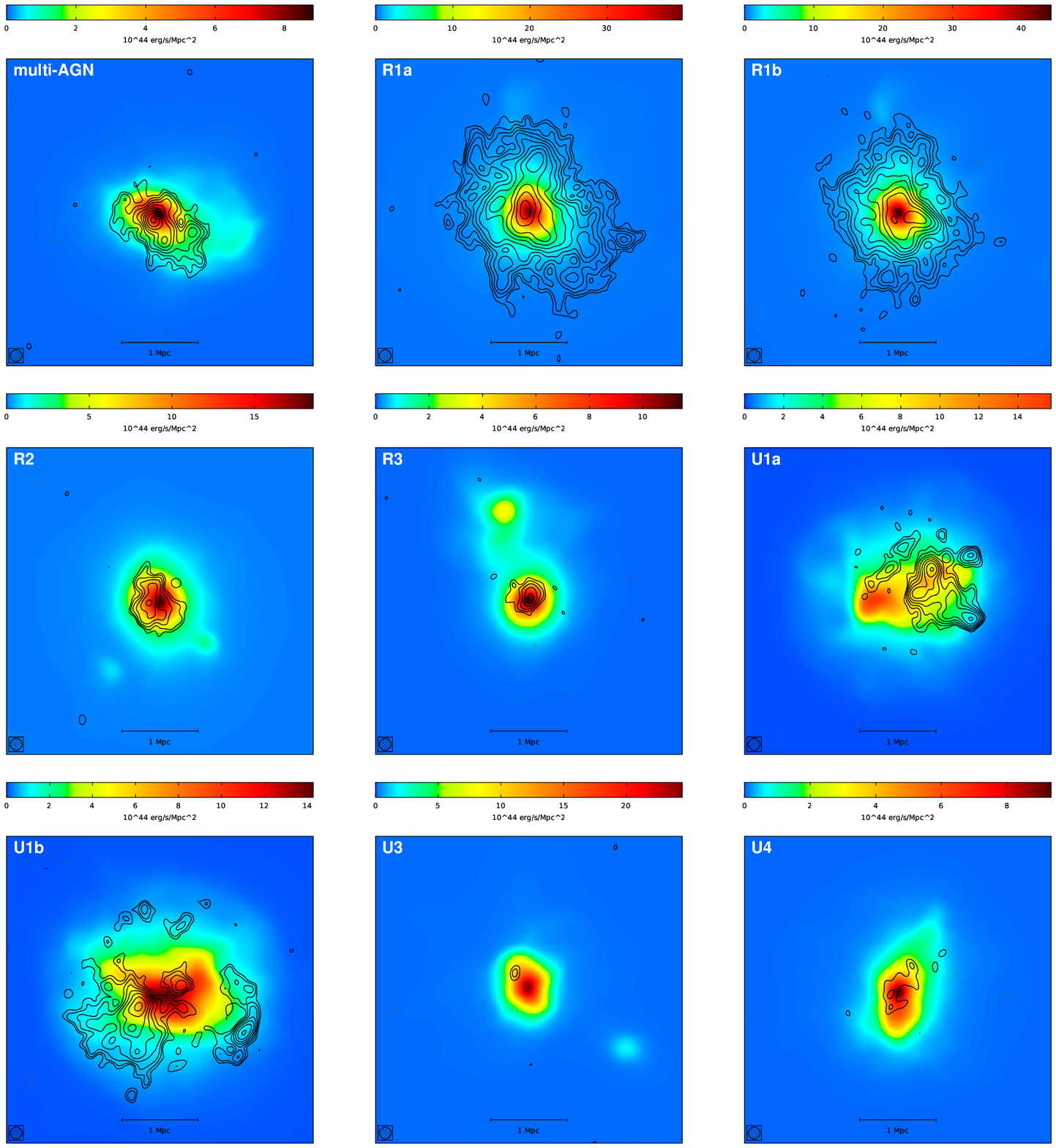,height=0.8\textheight} 
\end{center}
 \caption{Simulated clusters: multi-AGN, R1a, R1b, R2, R3, U1a, U1b, U3, and U4.
Radio halo emissions mapped to sky and filtered like 
typical VLA observations in D configuration with an integration 
time of 2 hours per cluster. 
The radio images have an angular resolution of 50$"$ FWHM. The contour levels 
start at 0.3 mJy/beam (3$\sigma$) and increase by $\sqrt2$. 
Radio halo contour levels are overlaid to 
the cluster X-ray emission in the 0.1-2.4 keV band smoothed at 45$"$ 
resolution to mimic a ROSAT PSPC observation. The radio emission from U3 is barely detectable, 
but we consider it as undetected since in a real image it would be indistinguishable from a background radio source.
   \label{fig:halos}}
\end{figure}

\begin{figure}
\begin{center}
\epsfig{file=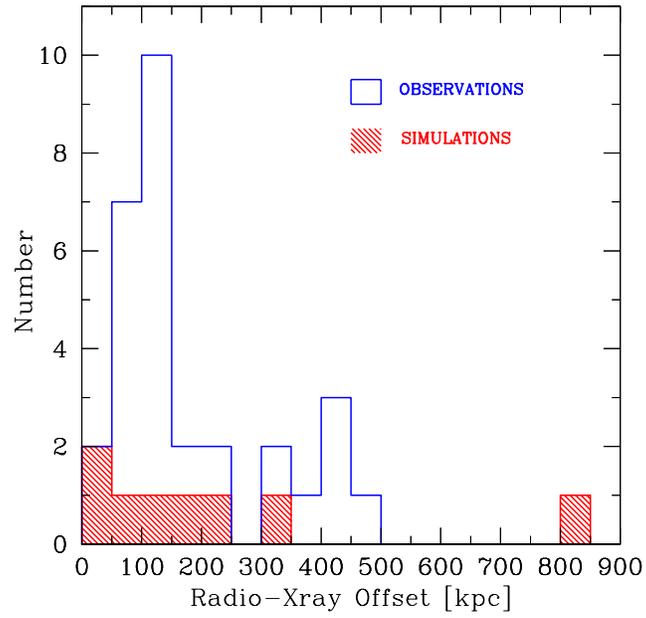,height=0.4\textheight} 
\end{center}
 \caption{Offset between the radio and X-ray peaks for a sample
of radio halos (from Govoni et al. 2012), compared with simulations. 
    \label{fig:offset1}}
\end{figure}

First we compare the offset between the X-ray and the radio peaks \citep[see][]{Feretti11}. This is a quite easy observable to
analyze and it can be related to the dynamic state of the ICM. In particular, it could be expected that unrelaxed clusters 
with ongoing mergers will show larger offsets with respect to relaxed clusters. Moreover, this offset could be directly 
 related to the magnetic field power spectrum on large scales \citep{Vacca10}. Very recently, \citet{Govoni12}
 measured this offset for a sample of  galaxy clusters containing radio halos by analyzing VLA and ROSAT data. 
In Figure \ref{fig:halos} we show the radio contours of the detected synthetic halos overlaid on the simulated cluster 
X-ray emission in the 0.1-2.4 keV band. We smoothed the X-ray emission to a resolution of 45$"$ 
 in order to mimic a ROSAT PSPC observation. In this way we can compute the offset between
 the X-ray and the radio peaks in a similar way as in most data of galaxy clusters hosting radio halos.
We note that the synthetic radio halos can be quite asymmetric 
with respect to the X-ray gas distribution.
In Figure \ref{fig:offset1} we compare the offset between the radio and X-ray peaks for a sample
of radio halos \citep{Govoni12} with simulations. 
The offset measured in the synthetic radio images are on the range of the observed ones, although a 
detailed comparison of the two distributions would require a much larger statistics for both simulations and data.
We note that the offset is comparatively small for the multi-AGN injection and for the relaxed clusters R1a, R1b, R2 and R3.
On the other hand, the offset is particularly large for the unrelaxed clusters U1a and U1b, as expected. In fact, the
 offset is so extreme for these systems that they could be even classified as cluster roundish relics according to
 \citet{Feretti12}.

\subsubsection{Comparison with Observed $P_{\rm 1.4 GHz}$--$L_X$ and $P_{\rm 1.4 GHz}$--$LLS$ Relations}
\begin{figure}
\begin{center}
\epsfig{file=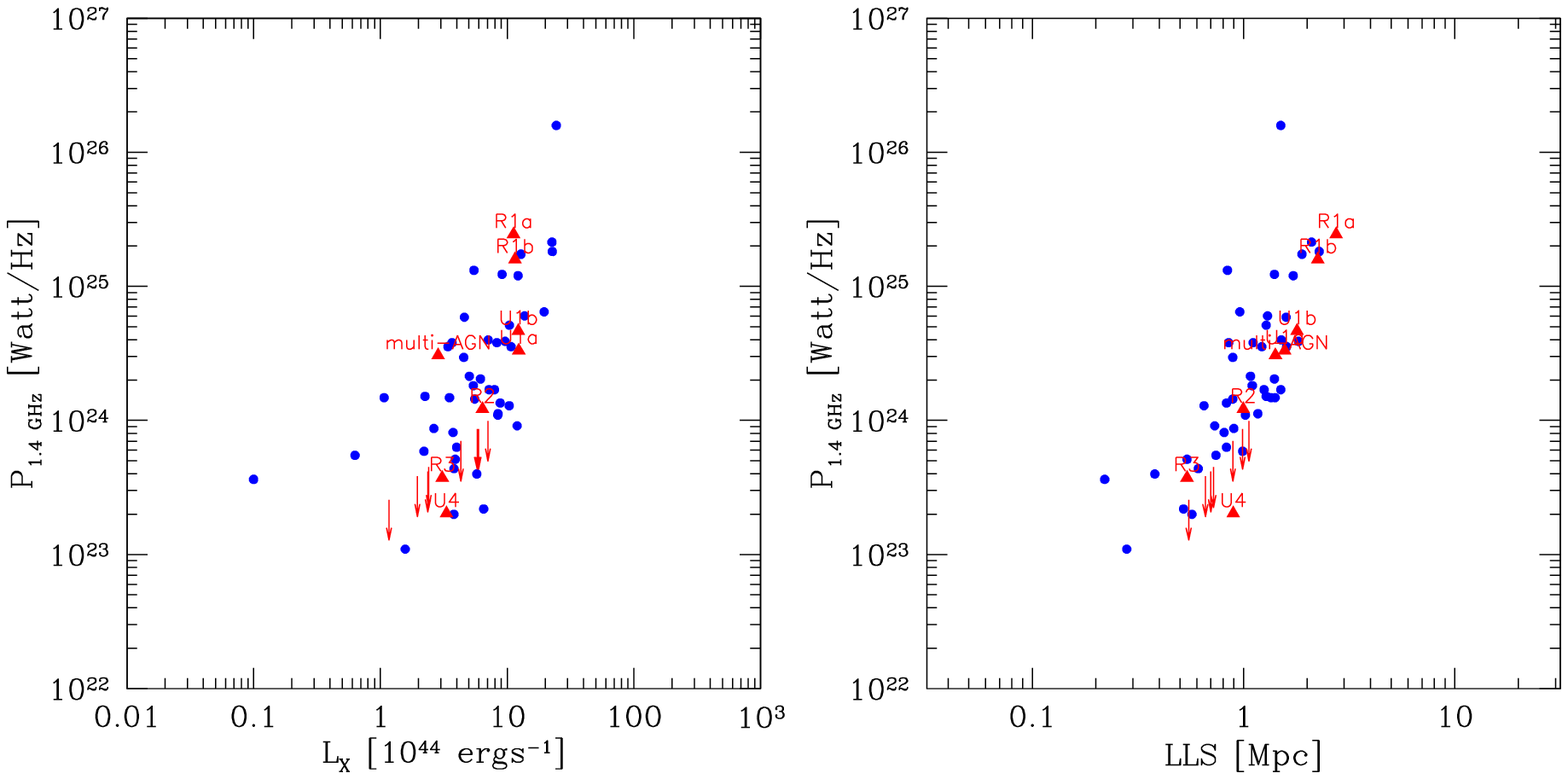,height=0.4\textheight} 
\end{center}
 \caption{
Left panel: radio power of radio halos at 1.4 GHz versus
the cluster X-ray luminosity in the 0.1$-$2.4 keV band. Right panel: radio power of halos at 1.4 GHz versus 
their largest linear size (LLS) measured at the same frequency.
Blue dots are observed clusters, 
red triangles are simulated clusters, while arrows indicate upper limits on the radio power of simulated clusters. The LLS are measured from the 3$\sigma$ isophote on the halo images.
The data are taken from the recent compilations by Feretti et al. (2012) 
and Govoni et al. (2012). 
    \label{fig:correlazioni}}
\end{figure}

In Figure \ref{fig:correlazioni} we plot the radio 
power calculated at 1.4 GHz versus the cluster X-ray
luminosity in the 0.1$-$2.4 keV band and versus the
Largest Linear Size.
Blue dots refer to the data published in the literature
\citep[taken from][]{Feretti12,Govoni12}, 
while red triangles refer to the simulations.
In the simulations $L_X$ has been calculated within a circle 
of 1 Mpc in radius, while $P_{\rm 1.4 GHz}$ and 
$LLS$ have been obtained by considering the radio halo emission
up to the 3$\sigma$ level.
Eight out of 16 simulated clusters present a detectable radio halo under the 
hypothesis of equipartition condition. Among the ``radio loud'' clusters we have the
relaxed clusters R1a and R1b and the unrelaxed clusters U1a and U1b. 
These four clusters are also the most X-ray luminous systems of the simulation set and they all 
have about the same X-ray luminosity of $\sim 10^{45}$ erg/s, but are quite different in radio 
luminosity.
In fact, the diffuse radio emission in R1a and R1b is about one order of magnitude more
powerful than in U1a and U1b. The difference could be related to their different evolutionary
 histories. According to the classification in \citet{Xu11}, R-type clusters have accreted half of their actual 
masses by $z=0.5$ while U-type clusters gained more than half of their final masses from $z=0.5$ to $z=0$.
The decaying turbulence originated from the early mergers in R1a and R1b has had enough time to diffuse and amplify the
 magnetic field over a large portion of the cluster volume, which results in a powerful and extended radio halo at
 the cluster center. On the other hand, the merging process is still in progress for U1a and U1b, and consequently 
there has not been enough time to amplify the magnetic field, to the level found in R1a and R1b.
The synthetic radio halos in R1a, R1b, U1a, U1b, together with those in R2, R3, U4 agree with the observed correlations
between the radio power versus the cluster X-ray luminosity and between the radio power versus the radio halo size.
The synthetic radio halo in the case of multi-AGN magnetic field injections 
 lies perfectly in the relation between radio power versus the radio halo size, but is over-luminous in radio with respect to the clusters R3 and U4 which have a very similar X-ray luminosity. This may imply that clusters in
which the injection of seed magnetic field by AGNs has been particular efficient may also host radio halos with
 a power higher than average. We speculate that this is could be a possible explanation also for the few radio over-luminous outliers known so far in the literature: A523 \citep{Giovannini11}, A1213 \citep{Giovannini09}, and 0217+70 \citep{Brown11}.
For the remaining 8 simulated clusters whose radio emission is below the 
detection threshold we can only derive upper limits. 
In order to calculate the upper limit on total radio power $P_{\rm 1.4 GHz}$ for 
the undetected radio halos we must first obtain an estimate of their putative size.
On the basis of the  $L_{X}$--$LLS$ correlation (not shown), we extrapolated the putative radio 
halo size from its X-ray luminosity. The upper-limits on total radio power $P_{\rm 1.4 GHz}$ were then calculated with the assumption 
that the average surface brightness of the diffuse emission over the putative halo size 
is lower than the $3\sigma$ noise level of the corresponding radio image.
We note that the clusters with a non-detected radio halo emission are the fainter X-ray clusters.

In summary, the magnetic field models presented here 
seem to generate halos of the right power and size, under the 
hypothesis of energy equipartition. By reversing the argument, we can deduce that 
the central magnetic field strength and radial decline in our simulations 
combine to produce a volume average magnetic field that matches the 
equipartition magnetic field estimates for observed radio halos.

\section{Conclusions}

To demonstrate the effects of location and number of injections on the ICM magnetic fields,  
we study the magnetic field evolution of a galaxy cluster with magnetic fields injected by as many as 30 AGNs at redshift $z=2.5$.  The cluster is well magnetized at low redshifts, the final mean cluster magnetic fields are $\sim$ 1 $\mu$G at the cluster center and $\sim$ 0.2 $\mu$G at its virial radius. Though having wider spreading and higher volume filling, the magnetic field distribution of this cluster is not significantly different from those of our previously single AGN injection simulations.  The additional injected magnetic fields do not increase the final magnetic energy and field strength dramatically, but only increase the volume filling of the cluster, especially in the outer parts of the cluster.
 The total magnetic energy of this cluster is consistent with the scaling relation with the square of cluster mass in the high end. We also find smoother radial profiles of magnetic field strength after major mergers, likely due to the fast amplification of the high volume filled magnetic fields.

We produce VLA filtered synthetic Faraday rotation images from a set of simulated galaxy clusters and compare them with real 
observations. The radial profiles of the synthetic $|RM|$ and $\sigma_{RM}$ from simulations agree 
with the observation results from \citet{Clarke01} and \citet{Govoni10}. The $|RM|$ and $\sigma_{RM}$, just like magnetic field 
strengths, are high in large (hot) clusters, as suggested by observations \citep{Govoni10}. We also 
compare the $\sigma_{RM}$ - S$_{X}$ relation between synthetic results and observations, and show that they are in a broad 
agreement, although there is an indication that the simulated clusters could be slightly over-dense and less magnetized  (up to about a factor of two) with 
respect to those in the \citet{Govoni10} sample.

We generate synthetic radio halos by assuming the energy equipartition between magnetic 
fields and non-thermal electrons, and we compare the global properties of the mock radio halos with the real ones.
The offset between the radio and the X-ray peaks for the synthetic halos are in the range of the observed values
 recently measured in a sample of clusters hosting radio halos. In particular, we find the offset is small in relaxed clusters and 
large in unrelaxed clusters, suggesting that this indicator is strongly related to the dynamic state of these systems.
Furthermore, the synthetic radio halos agree with the observed correlations between the radio power versus the cluster X-ray 
luminosity and between the radio power versus the radio halo size.  

Therefore, we can draw the important conclusion that the combined effects of the central magnetic field strength and the radial decline in our simulations 
can produce a volume average magnetic field that matches the 
equipartition magnetic field estimates for observed radio halos and, at the same time, is able to explain the observed value 
of Faraday rotation measure of radio sources in clusters. This results confirm the findings by \citet{Murgia04} and \citet{Govoni06} 
that if the ICM magnetic fields fluctuate over a wide range of spatial scales and decline in strength with radius, it is possible to 
reconcile the equipartition magnetic fields in radio halos and the magnetic field derived from the RM in galaxy clusters.

\acknowledgments

H.X. and H.L. are supported by the LDRD and IGPP programs at LANL and by DOE/Office of Fusion Energy Science through CMSO. D.C. is supported by Advanced Simulation and Computing Program (ASC) and LANL which is operated by LANS, LLC for the NNSA of the U.S. DOE under Contract No. DE-AC52- 06NA25396. Computing resources were supplied by the National Science Foundation  by an allocation TG-MCA04N012 on Kraken at the National Institute of Computational Sciences.  ENZO+MHD is developed at the Laboratory for Computational Astrophysics, UCSD, with partial support from NSF grants AST-0708960 and AST-0808184 to M.L.N. F.G. and M.M. thank the LANL for the support and hospitality during the preparation of this work. R.C. is supported in part by NASA grant NNX11AI23G.

\appendix
\section{RM Images Filtering}
When comparing the expectation of numerical simulations 
of cluster Faraday rotation images with data one must take into account
of the instrumental filtering affecting the observed RM images.
The observed RM images of cluster radio galaxies considered in this work have 
been obtained through the well known $\lambda^{2}-$law relating the 
observed polarization position angle $\Psi$ at wavelength $\lambda=c/\nu$ to 
the intrinsic polarization angle $\Psi_0$ at a given position in the 
radio source:

\begin{equation}
\Psi(\lambda)=\Psi_0+RM\lambda^{2}/(1+z)^2.
\label{lambda2law}
\end{equation}

By observing the radio source at different frequencies it is possible to 
obtain both RM and $\Psi_0$ by a linear fit to Eq.\,\ref{lambda2law}. 
The approach is straightforward in principle but complicated in practice by a
number of observational issues. One is the beam depolarization. 
The spatial distribution of polarization angles is 
not uniform across the radio source both because of the intrinsic 
morphology of the source itself and because of the Faraday rotation. Indeed, 
the incoherent sum of the disordered polarization vectors inside the observing 
beam results in an attenuation of the polarized signal. Another effect to 
consider is the bandwidth depolarization. The Faraday rotation of the
polarization angle inside the observing frequency bandwidth leads again to an 
attenuation of the signal. Finally, another difficulty to face is related to
the n-$\pi$ ambiguity of the observed polarization angles which must be 
resolved by the fitting algorithm. All these effects are 
more prominent at low frequencies and in general lead to non-linear 
distortions of the $\lambda^{2}-$law which are fairly a bit more complicated
than the deviations induced by a simple propagation of the error measurements
 from the polarized intensity images. The consequent error on the observed RM 
is difficult to quantify analytically.

For all these reasons, we filtered the simulated Faraday rotation images before  
comparing them to data. The filtering is performed with the FARADAY code 
and consists of the following steps:
\begin{itemize}
\item[i)] {first, we created two mock images of $4096\times 4096$ pixel in size 
for the distribution of intrinsic polarization intensity and angle at a
 resolution eight times finer than that of the simulated Faraday rotation images (i.e. $\simeq 1.3$ kpc). To mimic the
 intrinsic variations seen in real sources, the polarized 
intensity and angle distributions are created randomly with Gaussian 
fluctuations on spatial scales smaller than 100 kpc.}

\item[ii)] {the simulated RM images are gridded to the size of the intrinsic polarization 
images and we employed Eq.\,\ref{lambda2law} to rotate 
the intrinsic polarization vectors at frequencies of 8465, 8085, 4935, 4735, and 
4535\,MHz. This is the frequency setup used by \citet{Bonafede10} 
for the RM study of the Coma cluster, but it can be considered representative 
also of most other similar studies based on VLA observations;}

\item[iii)] {we produced synthetic polarization images of Stokes 
parameter $U(\nu)$ and $Q(\nu)$ at each frequency by adding a Gaussian 
noise such that we have an average signal-to-noise ratio of  $\langle P/\sigma_{P} \rangle =5$ for the 
polarization intensity a 8465 MHz. The polarized intensity at 
lower frequencies is scaled with a spectral index of 0.8, 
i.e. $P(\nu)=P_{8465}(\nu/8465)^{-0.8}$, so that the signal-to-noise ratio is slightly higher at
 lower frequencies;}

\item[iv)] { to simulate the beam depolarization the synthetic  $U(\nu)$ and $Q(\nu)$ images are convolved with 
a FWHM beam of $4\times 4$ pixels (which corresponds to about $3\times 3$ arcsec  at $z=0.1$). Moreover, to simulate
 bandwidth depolarization, the synthetic polarization images are averaged over a bandwidth of 50\,MHz centered on each frequency;}

\item[v)] {finally, the synthetic $U(\nu)$ and $Q(\nu)$ images have been gridded to the 
original size and resolution of the simulated Faraday rotation image and then analyzed
as if they were real polarization observations.
The synthetic RM images are created pixel by pixel by fitting the ``observed''
polarization angle versus the squared wavelength
for all the frequencies. To reduce the problems associated with
n-$\pi$ ambiguities, the fitting algorithm in FARADAY can perform a sequence of
improvement iterations. In the first iteration, only a subset of
high signal-to-noise pixels is considered. In the successive iterations,
lower signal-to-noise pixels are gradually included and the
information from the previous iteration is used to assist the fit of
the $\lambda^2$-law. In the procedure we clipped all pixes where error on the polarization
 angle at any frequency was greater than $15\degr$ and those where the reduced $\chi^2$ of the fit was worse than 10, as usually done with real data.}

\end{itemize}

\begin{figure*}
\begin{center}
\includegraphics[width=12cm]{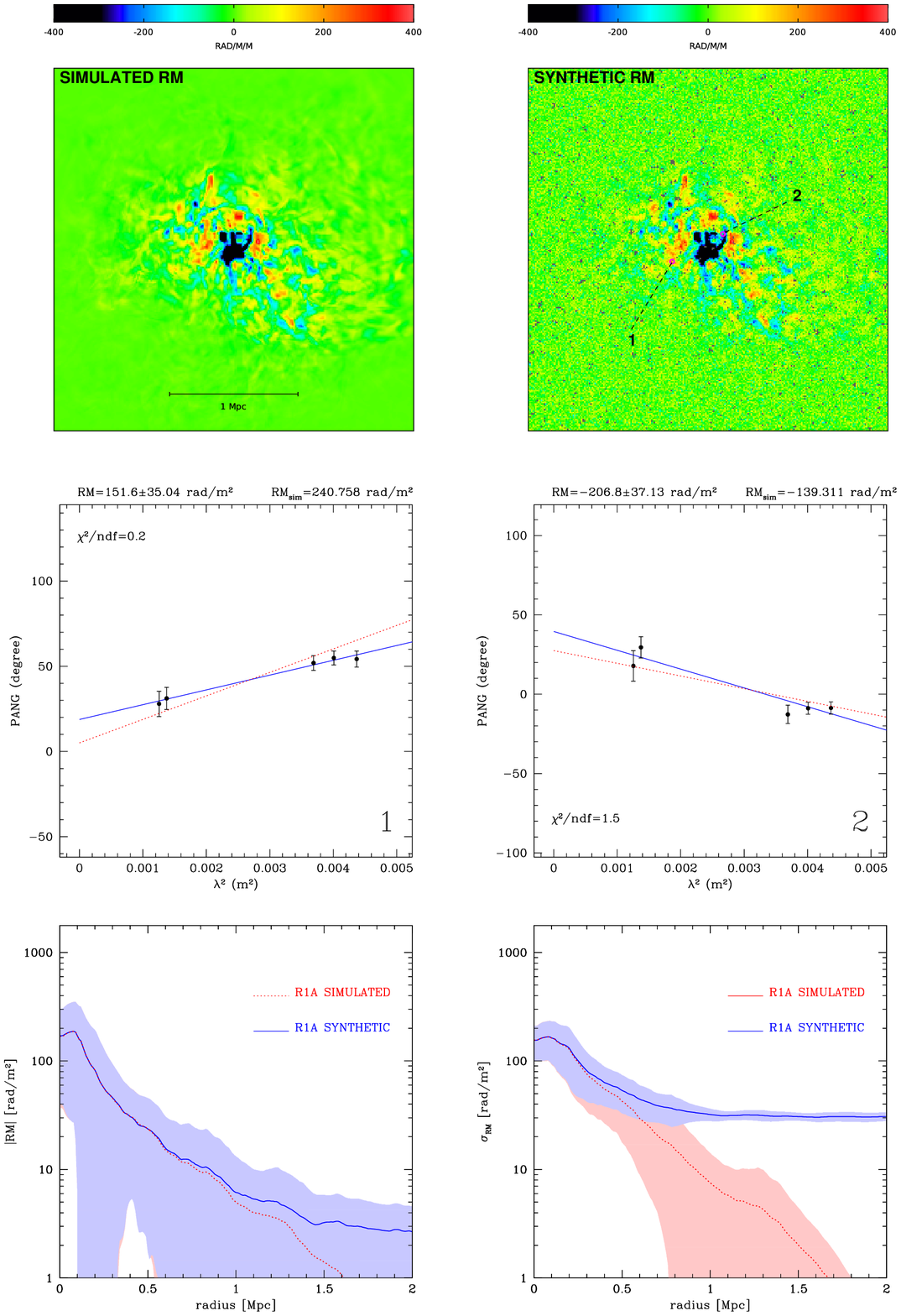}
\caption{Simulated and synthetic RM images of multi-AGN injection cluster at $z=0$ are shown in the top left and right panels, respectively. Middle panels show the fit of the $\lambda^{2}$-law (solid lines) at the two locations indicated in the synthetic RM image. The dotted lines correspond to the trend expected without instrumental filtering. The value of the unfiltered Faraday rotation value ($RM_{sim}$)  is also reported in each plot. Bottom left and right panel show the simulated and synthetic  $|RM|$ and $\sigma_{RM}$ profiles, respectively. The profiles represent an exponentially smoothed radial average of the RM statistics calculated in boxes of 
$100\times 100$ kpc$^2$. The shaded regions represent the rms scatter around the average profiles.}
\label{figa1}
\end{center}
\end{figure*}

A comparison of simulated and synthetic RM images is shown in the top panels of
 Figure \ref{figa1} for the case of the multi-AGN injection simulation at $z=0$. The instrumental 
filtering results in a RM noise of about 30 rad/m$^2$.  
In the central region of the cluster, where the RM signal is strong, the two images are 
 very similar, although the fit failed in a few isolated pixels (gray pixels). 
However, in the peripheral regions of the cluster the RM signal is too weak to be 
detected with this observational setup which is based on
relatively high-frequencies. Middle panels of  Figure \ref{figa1} show 
the fit of the $\lambda^{2}$-law (solid lines) at the two different locations indicated in the 
synthetic RM image. The dotted lines correspond to the trend expected without instrumental filtering.
Although the simulated and synthetic RM can be different from pixel to pixel, their statistical properties
 are quite stable. In the bottom panels of  Figure \ref{figa1} we show a comparison of the simulated and synthetic
 RM profiles of $|RM|$ and $\sigma_{RM}$ as a function of the distance from the cluster center. The RM statistics
 is calculated over boxes of $100\times100$ kpc$^2$ in size, to reproduce the typical size
of observed RM images of cluster radio sources. The profiles shown in Fig.\ref{figa1} represent
 a radial average obtained by exponentially smoothing with a length-scale of 50 kpc 
the $|RM|$ and $\sigma_{RM}$ calculated in all the boxes. We note that the simulated and synthetic profiles of $|RM|$ match each other, this means that the instrumental filtering has a minor impact on this 
 indicator. While the flattening of the synthetic $\sigma_{RM}$ profile at large radii is the 
most relevant effect introduced by the instrumental filtering.
The shaded regions represent the rms scatter around the average profiles. This scatter is not originated by instrumental errors but rather represents an intrinsic characteristic of the intra-cluster magnetic fields. In particular we note that the $|RM|$ is intrinsically much more 
scattered than $\sigma_{RM}$. This is due to the fluctuations of the intra-cluster magnetic field on scales larger than 100 kpc, i.e. the size of the boxes in which the RM statistics is calculated.

\begin{figure*}
\begin{center}
\includegraphics[width=12cm]{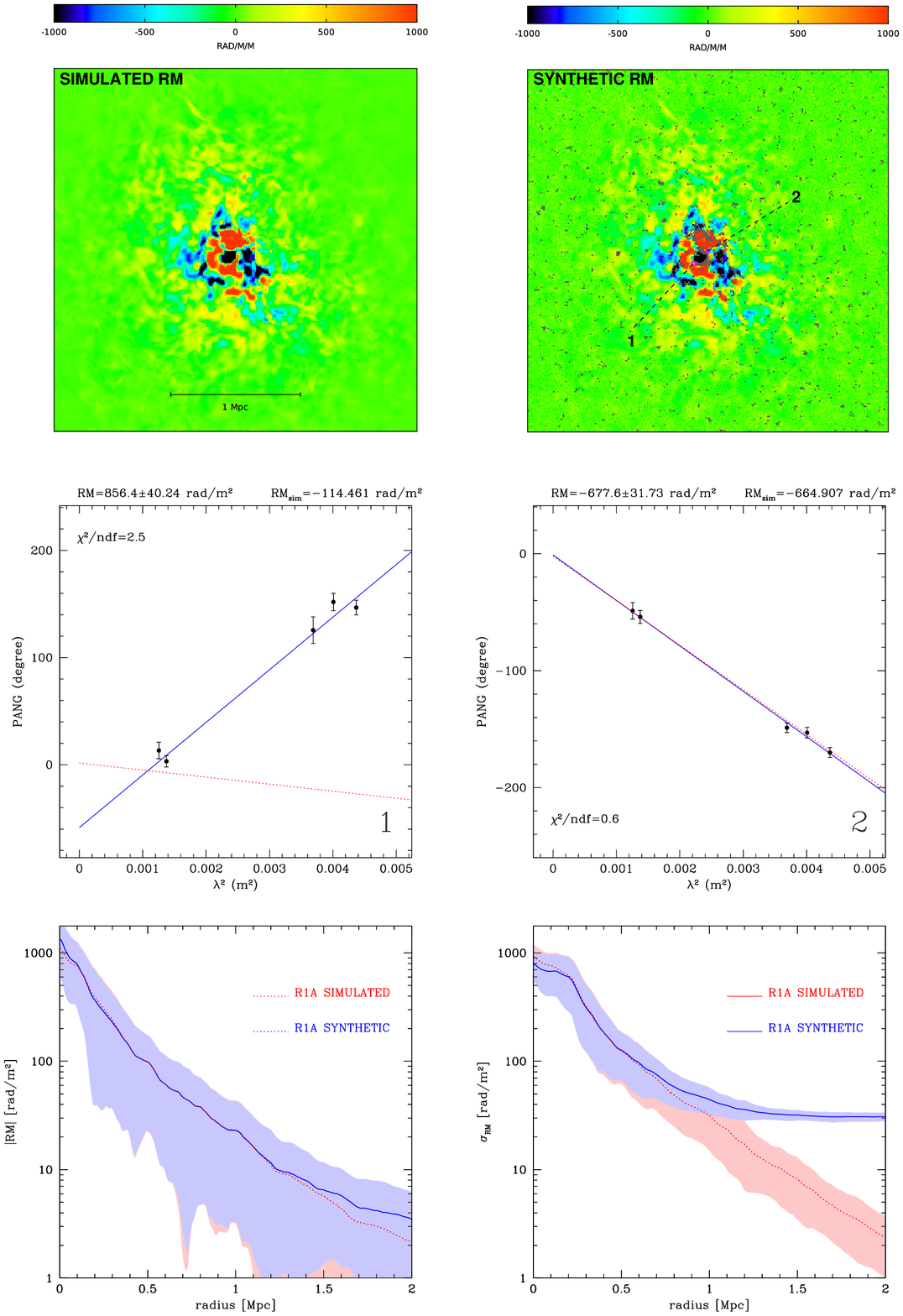}
\caption{Same as Figure \ref{figa1} but for the simulated cluster R1a. The stronger Faraday rotation gradients in this cluster causes large depolarization regions in the center (gray pixels). Close to these regions the fit of the $\lambda^2$-law can be troublesome, as shown in the middle-left panel. However, the radial profiles of the simulated and synthetic  $|RM|$ and $\sigma_{RM}$ shown in the bottom panels are in a remarkable agreement. The shaded regions represent the rms scatter around the average profiles.}
\label{figa2}
\end{center}
\end{figure*}

In Figure \ref{figa2} we show the comparison of simulated and the synthetic RM images for the case 
of the simulated cluster R1a. The stronger Faraday rotation gradients in this cluster causes large depolarization regions in the center. Close to these regions the fit of the $\lambda^2$-law can be troublesome, as shown in 
the middle-left panel. Here, we show the results for a pixel located at the interface of two RM cells of opposite signs. The beam and bandwidth depolarization caused by the strong RM gradient results in a deviation from the original $\lambda^2$-law (dotted red line). The fitting algorithm 
 then fails to solve for the correct n-$\pi$ ambiguities and the resulting $RM$ and $\Psi_0$ are completely wrong (continuous blue line and dots).

However, the radial profiles of the simulated and synthetic  $|RM|$ and $\sigma_{RM}$ shown in the bottom panels are in a remarkable agreement.

\section{Radio Halo Images Filtering}

As a consequence of their low surface brightness,
radio halos are generally studied through interferometric images
at a relatively low spatial resolution.
Thus, although radio halos may have an intrinsic filamentary structure,
the details of their morphology are often very poorly seen when 
observed with typical resolutions of $40-60\arcsec$.
In addition, interferometers filter out structures larger than 
the angular size corresponding to their shortest spacing, 
thus a loss of flux density is expected for radio halos with 
an extended angular size.
Indeed, in order to compare the simulated 
radio halos to the observations we produced synthetic radio halo
images which include the effects of the instrumental filtering.

In particular, we performed the following steps:
\begin{itemize}
\item[i)] {
first we performed mock radio halo images at 1.4 GHz 
with a bandwidth of 25 MHz by integrating the synchrotron emissivity
at each point of the computation grid along the line-of-sight.
}

\item[ii)] {these full resolution images are mapped as they would appear 
on the sky at a redshift z=0.2, which is the average distance of
the known radio halos. We also k-corrected their surface brightness
for the cosmological dimming;
}

\item[iii)] {
we filtered the resulting images in the Astronomical Image Processing System\footnote{AIPS is produced and maintained by the National Radio Astronomy
Observatory, a facility of the National Science Foundation
operated under cooperative agreement by Associated Universities, Inc.} 
with the set-up which has been typically used in many of the
 pointed interferometric observations of radio halos reported in literature. 
In particular, we converted the images in the expected visibility data 
for a 2 hours time-on-source observation with the VLA in D configuration; 
}

\item[iv)] {
the synthetic visibility data-set was then
imaged in AIPS as a real observation. At the end,
we obtained a filtered synthetic image of the radio halo 
characterized not only by the proper noise and spatial resolution, 
but also including the subtle effects induced by the missing short-spacing
 and by the imaging algorithm.
}
\end{itemize}

\begin{figure}
\begin{center}
\includegraphics[width=16cm]{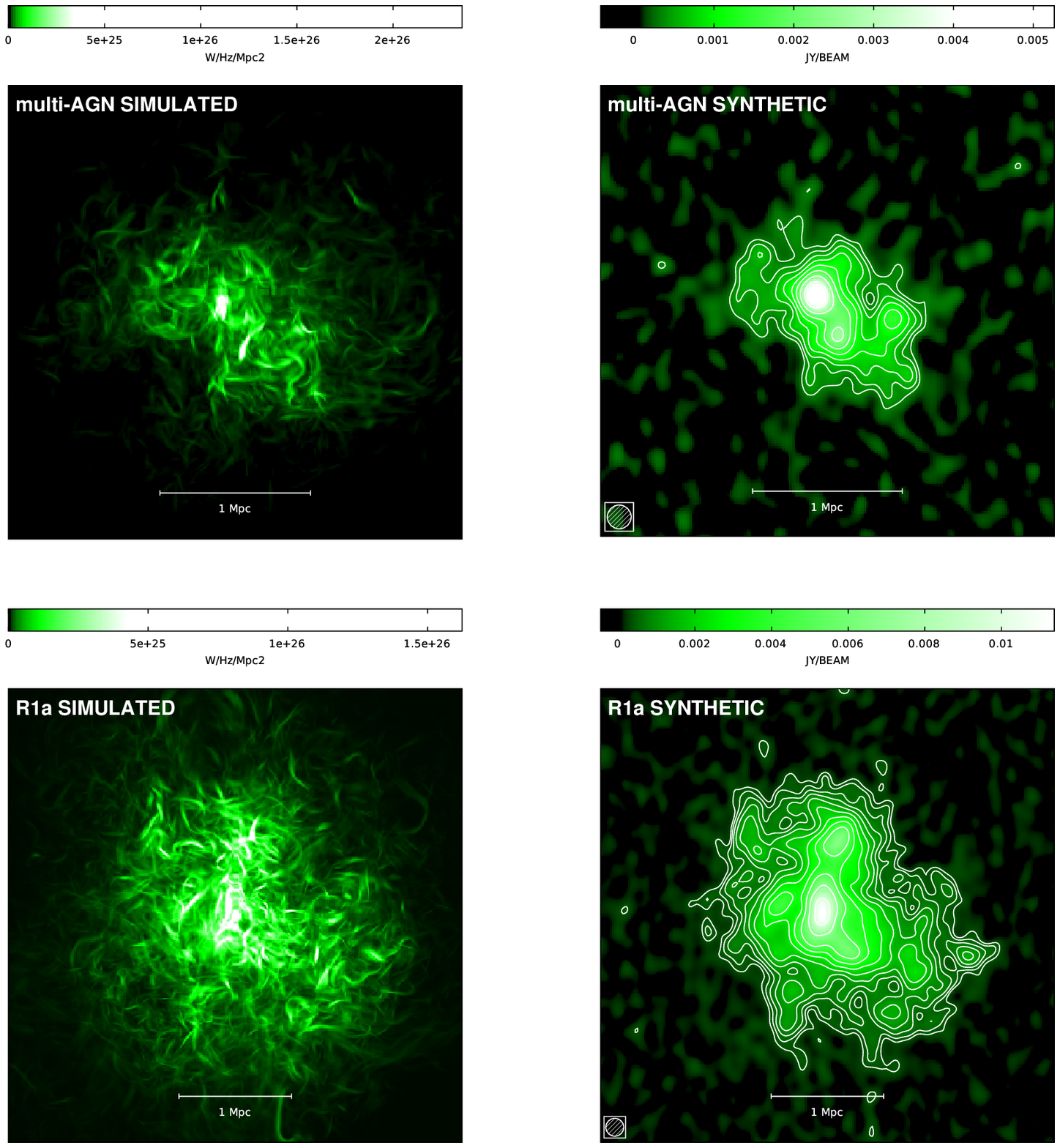}
\caption{Full resolution total intensity radio image at 1.4 GHz,
obtained from the simulated cluster with multi-AGN magnetic 
field injections at redshift $z=0$ (top-left panel) and the simulated cluster R1a (bottom-left panel).
The right panels show the synthetic radio halo emission mapped to sky and filtered 
like a typical VLA observation in D configuration 
with an integration time of 2 hours. The synthetic images have a resolution of 50$"$ FWHM beam. 
The contour levels start at 0.3 mJy/beam (3$\sigma$) and increase by $\sqrt2$. 
   \label{fig:sim30i}}
\end{center}
\end{figure}

Examples of the radio halo image filtering are presented in 
Figure \ref{fig:sim30i} for the simulated cluster 
with multi-AGN magnetic field injections at $z=0$ and for the simulated cluster R1a. 

The full resolution radio halo images are shown 
on the left column panels of Figure \ref{fig:sim30i}. 
In these images we can appreciate the full extension of the radio
halo emission and also the fine details of its intrinsic filamentary structure.

The synthetic radio halo images after the filtering process are shown
on the right column panels of Figure \ref{fig:sim30i}. 
The synthetic radio images have an angular resolution of 50$''$ FWHM  
and have a noise level of 0.1 mJy/beam (1$\sigma$). Most of the fine
 details of the halo structure are not visible anymore and also  
a significant fraction of the radio halo results well below the detection
 threshold, especially for the intrinsically fainter radio halo of
 multi-AGN magnetic field injections simulation.

\bibliographystyle{apj}
%\bibliography{ms}

\clearpage

\end{document}